# Polariton condensation phase diagram in wide bandgap planar microcavities: GaN versus ZnO

O. Jamadi[1,2], F. Réveret[1,2], E. Mallet[1,2], P. Disseix[1,2], F. Médard[1,2], M. Mihailovic[1,2], D. Solnyshkov[1,2], G. Malpuech[1,2], J. Leymarie[1,2], S. Bouchoule[3], X. Lafosse[3], F. Li[4], M. Leroux[4], F. Semond[4], and J. Zuniga-Perez[4]

[1]Clermont Université, Institut Pascal (IP), BP 10448, F-63000 Clermont-Ferrand, France
[2]CNRS, UMR 6602, IP, F-63171 Aubière, France
[3]LPN-CNRS, F-91460 Marcoussis, France
[4]CRHEA-CNRS, F-06560 Valbonne, France

**Abstract**

GaN and ZnO microcavities have been grown on patterned silicon substrate. Thanks to a common platform these microcavities share similar photonic properties with large quality factors and low photonic disorder which gives the possibility to determine the optimal spot diameter and to realize a complete comparative phase diagram study. Both systems have been investigated under the same experimental condition. Experimental results are well reproduced by simulation using Boltzmann equations. Lower polariton lasing threshold has been measured at low temperature in the ZnO microcavity as expected due to a larger Rabi splitting. However the threshold is strongly impacted by LO phonons through phonon-assisted polariton relaxation. We observe and discuss this effect as a function of temperature and detuning. Finally the polariton lasing threshold at room temperature is quite similar in both microcavities. This study highlights polariton relaxation mechanism and their importance for threshold optimization.

**Introduction**

Polaritons are bosonic quasiparticles arising from the strong coupling between excitons and photons [1]. When created within an optical microcavity, their shared light-and-matter properties can be easily tuned [2,3], which is of great interest to realize optoelectronic devices such as a polariton laser [4]. In contrast to a conventional laser, the polariton laser operation does not require population inversion. It is based on the stimulated polariton scattering by final state occupancy, which results in the formation of a macroscopic population of polaritons in the ground state. A coherent bright emission of the condensate [5,6] is then observed, with a reduced threshold [7] as compared with an ordinary laser. The first observation of polariton lasing was reported in a CdTe planar microcavity (MC) by Dang et al. [8] and a few years later, the Bose-Einstein condensation was obtained with polaritons up to 20 K [9]. At the same time, non-linear polaritonic emission was also observed in GaAs planar MC [7,10]. Up to now, many studies have reported phenomena based on the physics of polaritons and their condensates such as superfluidity [11], topological defects [12], parametric amplification, [13] and their possible applications, such as low-threshold lasers [14,15], optical switches and transistors [16–18], diodes [19], interferometers [20], and optical routers [21,22].

The high crystalline quality of epitaxially-grown GaAs structures enabled the observation of many fundamental effects, but its small excitonic binding energy is not suitable for room-temperature (RT) polariton devices. On the contrary, the robustness of excitons in GaN and ZnO at RT has led to an increasing interest for these materials, especially after the first observation at RT of the strong coupling regime (SCR) [23], and of polariton lasing in bulk- [24] and quantum well-based [25] GaN MCs elaborated on sapphire substrates. Concerning ZnO MCs, the SCR at RT has been reported more recently [26–29], followed by polariton lasing at 120K [30], then up to 250 K [31], and finally at RT [32,33]. The difficulty to increase the number of pairs of the distributed Bragg reflectors (DBRs) due to the lattice and thermal expansion coefficient mismatch between silicon and nitrides prevented the achievement of polariton lasing in such samples [34,35]. However, recently, the stacking of a large number of AlN/AlGaN pairs without the formation of cracks has been realized thanks to the use of mesa-patterned silicon substrates that allow strain relaxation by the lateral free surfaces. Optical cavities with large cavity quality factors in the near UV-range can be grown on silicon with high reproducibility. This progress has led to the first observation of polariton lasing at RT in bulk GaN and ZnO planar MCs on patterned silicon substrates [36]. The good optical and structural quality of the fabricated structures and their low photonic disorder now allows to carry out systematic studies of the polariton lasing threshold dependency.

The polariton lasing threshold depends on several parameters, which fall into two groups: kinetic ones (relaxation processes available and their efficiency, as well as particle lifetime) and thermodynamic ones (effective temperatures and polariton effective mass), which are determined by the detuning and the Rabi splitting. An in-depth knowledge of the influence of these parameters for a given material system is crucial to reduce the threshold in the aim to realize an optimum design for low consumption devices. Phase diagram (i.e. the T-dependent threshold-detuning characteristics) studies were already reported in planar MCs with GaN multi-quantum wells [37] and bulk ZnO [38] (as well as in CdTe and GaAs at lower temperatures). However, it is difficult from the analysis of these experimental data to compare quantitatively GaN and ZnO active layers, since both the photonic properties (in particular mirrors and active layer nature, QWs or bulk) of the studied MCs and the measurement conditions in these previous works were different.

In this study we compare rigorously experiments carried out under exactly the same conditions on two MCs on patterned Si substrates with similar photonic properties but different active bulk materials (GaN or ZnO). The first and second parts of this paper describe the samples and the calculation methods used. In the third section, the SCR is evidenced at RT in both systems whatever the detuning and excitation power. Thanks to the low cavities photonic disorder [36], it is also possible to study the influence of the spot size on the polariton laser threshold and the results are given and analyzed in the fourth section. Finally, a complete and detailed phase diagram is then determined for the two MCs from 10 K to 300 K over a wide detuning range by using the intentional thickness gradient of the active layer across the sample, which is 2'' diameter. Both systems are compared and confronted to numerical simulations based on Boltzmann equations [39] in order to gain further insight into the physics and efficiency of the relaxation processes. Finally, the large impact of the exciton-longitudinal optical (LO) phonon interaction on the polariton relaxation, which has been previously demonstrated in wide bandgap semiconductor microcavities [40,41], is addressed and its effect on lasing threshold reduction is analyzed showing that its efficiency is strongly dependent on the condensation/lasing regime in which the microcavity operates.

1. **Sample description and experimental setup**

Two MCs have been grown with similar photonic properties, but different active layers (GaN or ZnO). The bottom mirror is a 30-pair (or 30.5 pair) crack-free AlN/Al$_{0.2}$Ga$_{0.8}$N DBR directly grown on a patterned silicon substrate by molecular beam epitaxy (MBE). The size of the studied mesa varies from 100 to 500 µm. Both structures are covered with an 11-pair SiO$_2$/HfO$_2$ DBR. The only difference between the two cavities lies in the optical thickness of active layer: 3$\lambda$ of GaN (corresponding to a physical thickness of 414 nm) or 7$\lambda$/4 of ZnO (physical thickness of 290 nm), which were both epitaxially grown on the bottom DBR by MBE too. Further details on the fabrication are given elsewhere [36]. The thicknesses of the active layer have been chosen in order to obtain a Rabi splitting 2 times larger in ZnO.

These MCs are studied through micro-photoluminescence (µPL) experiments using the fourth harmonic (266 nm) of a Nd:YAG laser as the excitation source. The pulse duration is 400 ps, i.e. the excitation is quasi-continuous, and the repetition rate is 20 kHz. An UV-objective or an aspherical lens are used to obtain small excitation spot diameters of 3 and 8 µm respectively. Far-field emission measurements in reciprocal space are carried out by using a UV-objective with a numerical aperture of 0.5 giving access to a large range of emission angles (-10 to 50°). Samples are mounted inside a liquid helium circulation cryostat allowing measurements from 5 to 300 K.

2. **Theoretical model and simulations**

The linear optical properties of these MCs have been modeled as a classical harmonic oscillator coupled to light through Maxwell's equations. Due to the strong band-to-band absorption continuum, the upper polariton branch (UPB) cannot be detected experimentally in thick bulk active layers, typically above λ (~140 nm) in GaN [42] and λ/2 (~80 nm) in ZnO [43]. The Rabi splitting (Ω) is therefore determined from transfer matrix simulations using the recent accurate determination of the excitonic oscillator strengths [44,45]. When removing band-to-band absorption in the simulation in order to reveal the UPB, the Rabi splitting values are estimated to be equal to 80±10 and 160±10 meV for the GaN and ZnO MCs, respectively. Note that the Rabi splitting corresponding to the GaN MC is rather high compared to other structures [46–48] due to the use of a thick active layer, but close to the recently measured value of 84 meV in GaN MC with III-nitride air-gap DBRs [49]. In order to evaluate the excitonic and photonic weights of the cavity polaritons, a 2x2 coupled oscillator model [50] is used, from which the exciton-photon detuning (δ) is derived. The latter is defined as the difference between the energies of the bare cavity mode and the excitonic transition. A detuning equal to zero corresponds to polariton with identical excitonic and photonic fractions (50%); a positive (resp. negative) detuning corresponds to a larger excitonic (resp. photonic) fraction in the LPB. This model allows to simulate the polaritonic dispersion relation as a function of the emission angle or, equivalently, in-plane wavevector.

In order to calculate the phase diagram of polariton lasing we have performed simulations based on semi-classical Boltzmann equations for polaritons [5], taking into account polariton-phonon (both acoustic and optical) and polariton-polariton interactions. In the simulations we consider the populations $n_k$ of the states with wavevector $k$ in the LPB and in the excitonic reservoir. The evolution of these populations is given by:

$$\frac{dn_k}{dt} = P_k - \Gamma_k n_k - n_k \sum_{\mathbf{k'}} W_{k \to k'}(n_{k'} + 1) + (n_k + 1)\sum_{\mathbf{k'}} W_{k' \to k} n_{k'}$$

Here $P_k$ describes the non-resonant quasi-continuous pumping (particles appearing in high-energy states of the excitonic reservoir), $\Gamma_k$ describes the decay due to photon or exciton lifetimes (the temperature dependence of the last was taken into account), $W_{k \to k'}$ and $W_{k' \to k}$ are the scattering rates between the corresponding k and k' states. The polariton-polariton scattering rates depend on the current population of all states, and so need to be regularly updated during the calculation.

The calculation is carried out for 400 ps, which is the duration of the quasi-continuous pumping pulse. The population of the states close to $k = 0$ is integrated over this period of time, and the resulting "emission intensity" data are plotted as a function of pumping intensity, which allows to determine the threshold for given parameters. The calculations are then repeated over the whole range of detunings for the two materials, GaN and ZnO, in order to obtain the phase diagrams.

The material parameters (sound velocities, densities, exciton binding energies and Bohr radii, LO phonon energies) used in the calculations were the same as in Ref. [14] for GaN and Ref. [38] for ZnO. The values of the Rabi splitting were taken from the results of the optical calculations described above, and the photon lifetime was taken in agreement with the measured quality factor. In the absence of experimental data, the reservoir exciton lifetime was assumed to be 250, 100 and 50 ps at 20, 77 and 300 K correspondingly in both GaN and ZnO. This is one of the main parameters which increases the lasing threshold at high temperatures. Another fitting parameter was the excitonic inhomogeneous broadening, which was assumed to be 45 meV for both GaN and ZnO. This value is probably overestimated because of three reasons. First we take into account a single broadened excitonic resonance instead of two A and B excitons. So the broadening we use should be roughly increased with respect to the one of a single line by the value of the A-B splitting, which in total gives about 20 meV. The second reason is possibly linked by the necessity to compensate the underestimation of the scattering rates of the exciton-exciton and exciton-phonon interactions while converting the 2D problem to 1D using its cylindrical symmetry. Another cause of the underestimation of the scattering rates might be simply the crudeness of the numerical grid, which artificially increases its steepness, thus decreasing the efficiency of relaxation.

### 3. Analysis of the strong coupling regime at room temperature for various detunings

The SCR at RT was already observed and reported in these structures [36]. Here we will evidence that the SCR holds over the whole range of photonic fractions and excitation powers used in this study. Figures 1 and 2 show the polaritonic dispersion at RT in ZnO and GaN MCs for two different exciton-photon detuning values.

In the case of ZnO, below threshold the variation of the curvature of the lower polariton branch (LPB) observed in figure 1 (a and b) is a signature of the SCR. The photonic/excitonic fraction of the LPB can be easily chosen by moving the excitation spot from one mesa to another one along the sample's thickness gradient. The variation of the excitonic fraction of the LPB leads to the modification of the effective mass of the polariton. Experimentally it is directly evidenced through the deviation of the LPB curvature from a bare cavity dispersion (curve labeled C, dashed line) toward a pure excitonic dispersion (flat in this angle range due to the large excitonic effective mass). At the onset of polariton lasing (figure 1 c and d) we observe a sharp maximum of intensity corresponding to the polariton lasing emission, with a blueshift of around 10 meV [51], which is small compared to the Rabi splitting (6% of $\Omega$). The dispersions shown in Figures 1 (c) and (d) were measured at threshold in order to detect the emission of the LPB together with that of the condensate. It can be seen that the curvature of the LPB is not modified, attesting that the SCR still holds. Moreover, even for the most negative detuning accessible within the sample ($\delta$=-98 meV), the polariton laser line lies 40 meV below the bare cavity mode. Simulations carried out in the linear regime using the 4x4 coupled oscillator model (one exciton and three photonic modes C, B1 and B2) are in agreement with the experimental data.

The same measurements were performed on the GaN MC (see figure 2). The situation is somehow more complicated because of the presence of Bragg modes closer to the LPB. However, the SCR is evidenced by the change of the LPB curvature when the excitonic fraction is varied (figure 2a and b). A larger excitonic fraction results in a less pronounced curvature due to a larger polariton effective mass which confirmed the SCR, as previously observed [36]. Moreover, a curvature change between B1 and B3 can be observed which confirms that B1 is in strong coupling despite a quality factor much lower than the cavity mode. At threshold (Fig. 2d), as for the ZnO MC, SCR is preserved since the curvature of the LPB does not change and the polariton lasing takes place 20 meV below the bare cavity mode. The typical blueshift at threshold represent 10% of the Rabi splitting. For a more negative detuning the bare cavity mode becomes close to the LPB. However, polariton lasing is observed for power densities up to 200 W.cm$^{-2}$ at less negative detuning where the LPB is far from the bare cavity mode (Fig 2d). Thus, as Pth=150 W.cm$^{-2}$ in figure 2c, the SCR still holds even if the photonic fraction is high (85%).

To evidence that the SCR still holds even above threshold, we performed power dependent measurements on bulk GaN layer and on the half GaN MC (i.e. before the top DBR deposition). All the corresponding results are given in Annex (figures A1 and A2). The experimental data show that the transition from the luminescence from an exciton gas to an electron-hole plasma is not abrupt, but at least they enable to fix a limit for the input power (1.1 kW/cm²) below which we can state unambiguously that excitons still exist. In the present study polariton lasing is investigated well below this limit, in the range from a few W/cm² to 300 W/cm². Furthermore in the GaN MC investigated in this work, the condition requested to reach the SCR is fully satisfied since the Rabi splitting value is larger than the inhomogeneous excitonic broadening ($\Omega \gg \sigma$) [52,53]. In this sample the inhomogeneous excitonic broadening deduced from the fit of the half cavity reflectivity spectrum (figure A2b) is equal to 10±3 meV for A and B excitons.

4. **Influence of the spot size on polariton lasing threshold**

To investigate the MC photonic properties, µPL measurements were performed at RT for a large photonic fraction (82%). The energy of the LPB emission well below threshold and along a typical mesa is shown on figure 3a. While a maximum variation of 6 meV is observed for a displacement of 150 μm on a single mesa, a decrease of the energy variation down to 1 meV can be obtained in 50 μm wide areas. These measurements are in full agreement with the 2D cartography recorded on a similar structure [36]. Note that the average cavity quality factor on these samples is 2000 and could locally rise to 3550 (inset of figure 3a). These highly homogeneous areas for UV planar MC highlight the low photonic disorder that can be obtained with these GaN and ZnO MCs grown on patterned DBRs. Such structures are ideal to study the influence of the spot diameter on the polariton laser threshold (figure 3b). The average threshold (10 measurements per spot size) is measured on a same mesa at RT for a comparable photonic fraction (70-80%). For a small spot diameter, 3 μm, the high exciton density under the excitation spot induces a strong anti-trapping potential which ejects the created polaritons out of the pumped area due to polariton-polariton and polariton-carrier repulsive interactions [54]; this consequently increases the threshold ($P_{th}$). When the spot diameter is increased, the reduction of the repulsive potential gradient leads to a decrease of $P_{th}$ corresponding to an optimal spot size of 7-8 μm (figure 3b). Above 10 μm, the slight increase of $P_{th}$ is attributed to the small photonic disorder which degrades the quality factor. An excitation spot diameter of 8 μm corresponding to the best efficiency for the polariton condensation and this experimental condition was therefore fixed for the phase diagram studies.

5. **Comparison of phase diagrams of ZnO and GaN planar MCs**

In this section the influence of the detuning and the lattice temperature on the polariton lasing threshold is analyzed for ZnO and GaN MCs independently, and subsequently a comparison between these two systems is carried out. The experimental results are compared with simulations based on the semi-classical Boltzmann equations. The role of the LO phonons in the relaxation processes and their impact on the phase diagram is discussed.

A typical, the phase diagram of the ZnO MC measured at 100 K is reported in figure 4, evidencing the general features of the condensation phase diagram, with different regions being governed by the thermodynamics, the relaxation kinetics, or both [37].

(i) For large negative detunings, the polaritonic trap is deep (up to 177 meV), the polariton relaxation is slow, and the polariton lifetime, mainly photonic, is short. Polaritons cannot dissipate a sufficient amount of energy to reach the ground state only by scattering with acoustic and optical phonons. Thus, the pumping power has to be increased to enhance polariton-polariton interactions for accelerating the relaxation of the particles from the excitonic reservoir to the bottom of the LPB. As expected, the threshold increases when going towards more negative detunings. The polariton distribution is not in equilibrium and the threshold is governed by the kinetics of the relaxation processes.

(ii) For large positive detunings, the polariton trap is shallow, the polariton lifetime is always longer than the relaxation time; this leads to an efficient thermalization of the polariton gas, although its temperature can be different from that of the lattice [55]. The threshold behavior is then predicted by thermodynamics: one can expect an increase of the threshold versus detuning, as the polariton mass increases. Indeed, at larger detunings polariton interactions increase (due to the large excitonic part) and thermal escape of polaritons from the trap becomes significant, leading to a higher threshold. Between these two regimes (kinetic and thermodynamic), an optimal detuning ($\delta_{opt}$) where the threshold power is minimum exists. It corresponds to the trade-off between the polariton lifetime and the relaxation time [37,55]. It has already been observed in CdTe[55], GaAs[56], ZnO[40] and GaN[41] MCs, that polariton interactions with LO phonons can be an additional efficient relaxation process and contribute to a threshold reduction. In the following, we will clearly demonstrate that the efficiency of this LO phonon-assisted relaxation depends strongly on the detuning. In the thermodynamic regime, when the polaritonic trap depth – defined by the energy difference between the excitonic reservoir and the bottom of the LPB ($E_X$-$E_{LPB}$) – is equal to the energy of one LO phonon (this resonance is marked by a vertical dashed lined labeled 1LO in figure 4), LO phonons have no noticeable effect on the threshold. In this region of shallow traps polaritons have enough time to relax, even without the contribution of additional processes. However, for a negative detuning δ=-100 meV at the resonance with 2 LO phonons, a slight decrease of $P_{th}$ is observed, indicating that LO phonons contribute more efficiently to polariton scattering towards the ground sate. However, this process is not efficient for large negative detuning because the short polariton lifetime prevents their interaction with phonons. Actually, phase diagrams measured at different temperature show that the impact of LO phonons is the strongest when the resonance (either with one or two LO phonons) occurs in the trade-off zone, as presented below.

Complete phase diagrams measured from 10 to 300 K in GaN and ZnO MCs are reported on figure 5 and 6 respectively. The overall behavior corresponds very well to the qualitative description given above: each diagram presents a clear minimum between the thermodynamic and kinetic regions.

In the case of the GaN MC, an unexpected trend is observed for a detuning of -50±10 meV: the polariton lasing threshold decreases when going from 10 to 100 K (figures 5a). From a thermodynamic point of view, one would rather expect a threshold increase with temperature. This experimental behavior can be explained by the impact of LO phonon-assisted polariton relaxation on the threshold. Indeed, at 10 K the detuning corresponding to the LO phonon resonance with the bottom of the trap is strongly negative and the impact of LO phonons on the threshold is weak. Moreover, the kinetics becomes faster with increasing temperature. As a result, under negative detuning, the threshold could be lower at 77 K with respect to 10 K, because the kinetics controls the condensation. At 77 and 100 K, the LO phonon energy shifts towards positive detunings and enters the trade-off zone ($\tau_{pol}\sim\tau_{rel}$), where the impact of LO phonons on the threshold reduction is the most pronounced (figure 5a). Indeed, this relaxation mechanism results in a decrease of the relaxation time which extends the thermodynamic condensation region towards negative detunings.

In the low temperature range from 10 K to 220 K in the case of GaN and from 10 K to 150 K for ZnO and for detuning values typically larger than -20 meV (GaN) and 10 meV (ZnO), (figure 5 and 6), data merge into a single "curve". This feature clearly indicates that the polariton gas reaches a thermal self equilibrium at a temperature higher than the lattice temperature, which does not change significantly in this range of lattice temperatures. It has been verified that the lattice temperature is equal to the experimental setup temperature: power dependent measurements on the MC (not shown here) confirm that there is no heating of the lattice whatever the excitation power used, since we do not observe any energy shift of the phonon replica of free exciton. The relaxation time is shorter than the polariton lifetime but the kinetics is mostly driven by polariton-polariton interactions and, therefore, polaritons thermalize to an effective temperature $T_{eff}$ higher than the lattice one.

When the temperature increases above 150 K, we observe an increase of the threshold at $\delta_{opt}$ in both systems As for the ZnO MC, the lower energy of LO phonon (72 meV, compared to 91 meV in GaN). Concerning the ZnO MC, the lower LO phonon energy combined with both the accessible detuning range and the strong Rabi splitting, enables 1 LO- and 2 LO-assisted transitions to be in resonance with the polariton ground state within the investigated detuning range. At 10 K, the detuning leading to the 1 LO resonance is found within the trade-off zone and consequently the threshold is significantly reduced. At 77 and 100 K, the corresponding detuning shifts into the thermodynamic region and, as expected, no influence on threshold is observed. On the other side, the 2 LO resonance leads to a rather weak decrease of the threshold, since it occurs for rather negative detunings corresponding to the kinetic region. With increasing temperature, the optimal detuning shifts strongly towards more negative values, until the 2 LO resonance occurs in the trade-off area at RT. The threshold is consequently strongly reduced.

Figure 7 (a) and Fig. 7 (b) show the theoretical phase diagrams for GaN, and ZnO, respectively. Panel a) presents the phase diagram for GaN and panel b) for ZnO. Calculations were performed for 3 representative temperatures: 20, 77 and 300 K. One can see that the overall behavior of the lasing threshold is well reproduced.

The optimal detuning (black squares) together with the corresponding LO resonances as a function of the temperature are reported in figure 8 for both systems. The optimal detuning values correspond the average detuning of 10% of the lower measured threshold values. The dark blue area indicates the extension of the trade-off region and the light blue zone corresponds to error bars of this estimation. For GaN, between 77 and 150 K, the optimal detuning is very close to the 1 LO resonance, evidencing that the latter has a strong impact on the threshold value. For higher temperatures and more negative detunings, the 1 LO resonance goes out from the trade-off area and has no longer influence on the optimal detuning position. For ZnO, both one and two LO phonons can be involved in the polariton relaxation. From 10 to 300 K, the optimal detuning shifts continuously from 7 meV to -85 meV. Note that both 1 LO and 2 LO resonances (respectively at low temperature and RT) occur in the trade-off zone and, as previously observed in a ZnO MC with large Rabi splitting [38], have a strong impact on the threshold and on the energy of the optimal detuning value.

In order to compare both systems (GaN versus ZnO), experimental thresholds recorded at 10 K are reported on figure 9a as a function of the polaritonic trap depth. ZnO, with a higher Rabi splitting, has a lower polariton lasing threshold as expected [57,58]. At 10 K, for $\delta_{opt}$, the threshold is three times smaller in ZnO than in GaN. Note that the Rabi splitting of the ZnO MC is twice larger that of the GaN MC. It can be seen that such a large Rabi splitting enables to obtain polariton lasing even for deep polariton traps while keeping a non negligible excitonic fraction. At 10 K, the polariton trap accessible in this sample varies from 185 meV to 51 meV which corresponds to excitonic fractions of 16 and 71% respectively. Recently, it has been demonstrated that polariton condensation could be achieved in a thick hybrid ZnO MC at low temperature up to an extreme excitonic fraction of 96 %, thanks to a huge Rabi splitting value of 240 meV [38]. For a given polaritonic trap depth, the excitonic fractions related to the GaN MC are two to three times smaller compared to those corresponding to the ZnO MC. For the GaN MC sample, it is possible at 10 K to achieve polariton lasing with a low excitonic fraction (7%) at $E_X-E_{LPB}$=145 meV. This excitonic fraction value is increased to 65% ($E_X-E_{LPB}$=29 meV, $\delta$=25 meV on figure 5b) when the temperature is increased up to 220 K (kT=19 meV). Above this value, the condensation threshold increases so much that it prevents the observation of polariton lasing.

The threshold values at RT are compared on figure 9b versus the detuning. Showing both phase diagrams on the same figure allows to compare the two materials in view of applications at this temperature. A regime thermodynamically driven by polariton effective temperature and masses takes place for both system for a detuning values from -60 to 10 meV. This leads to close thresholds obtained in GaN and ZnO MCs with only a slightly lower threshold observed for ZnO at the 2 LO resonance.

The emission intensity is larger in the ZnO MC whatever the optical pumping intensity, as reported in figure A3. We observe intensities around 10 times higher just below and above threshold. They can be explained by the oscillator strength, which is 4-6 times larger in ZnO than in GaN [45], and the leakage of photons through the Bragg mode, which is more pronounced in GaN as observed on figure 2. The power dependence in ZnO well below threshold (first $P^1$ and then $P^2$) is in full agreement with the polariton assisted relaxation with LO phonon previously reported [40].

**Conclusion**

Both GaN and ZnO planar microcavities having similar photonic properties have been studied. The strong coupling regime was clearly observed whatever the excitonic fraction in the whole range of optical pumping powers used in this work. The analysis of the influence of the excitation spot diameter on the polariton lasing threshold indicates an optimal spot size of 8 µm. A complete phase diagram has been measured from 10 to 300 K in both systems. At low temperature, ZnO thanks to a larger Rabi splitting exhibits a lower polariton lasing threshold, as expected. The threshold decrease with temperature (from 10 K to 100 K) observed for the GaN microcavity is mainly due to the efficient polariton relaxation through LO phonons. At room temperature a regime thermodynamically driven by polariton effective temperature and masses leads to similar thresholds for both system with only a slightly lower threshold observed for ZnO at the 2 LO resonance. The strong influence of LO phonons, especially in the trade-off zone between thermodynamics and kinetics, is clearly observable for the two materials systems. In the case of ZnO, when the temperature is increased, the decreasing value of the optimal detuning becomes even compatible with a 2 LO phonon assisted relaxation at room temperature. Overall, this study highlights the polariton relaxation mechanisms in ZnO and GaN microcavities as function of temperature and detuning value. Their consideration should enable to design more efficient cavities displaying even lower polariton lasing thresholds.

**Acknowledgments**

We acknowledge support from GANEX (ANR-11-LABX-0014). GANEX belongs to the public funded 'Investissements d'Avenir' program managed by the French ANR agency.

**Annexes**

To confirm the observation of the SCR whatever the detuning, power-dependent experiments were performed on a high quality bulk GaN layer and on the half GaN microcavity (i.e. before the top DBR deposition). Concerning the experimental results obtained at 5 K on the GaN bulk layer (figure A1), the excitonic transitions are well detected in the spectrum at low excitation intensity (11 W/cm²) in agreement with previous study [59]. Peaks broaden when the excitation power is increased. Above 1.1 kW/cm², the A exciton ($X_A$) is no more detectable (red spectrum). As observed by Rossbach et al., the transition from an exciton gas to a so-called electron-hole plasma corresponding to the Mott transition, is not abrupt but rather continuous and does not depend on temperature [60]. Hence, we consider 1.1 kW/cm² as the lower limit below which excitons still exist. The determination of the Mott transition is beyond the scope of this study but one can notice that the optical pumping intensity used to reach polariton lasing is well below this low limit. The same measurements were performed on the corresponding GaN half MC, i.e. without the top DBR (figure A2 a-b). Reflectivity experiments combined with a low power PL measurement allow to identify $D^0X$, A and B excitonic transitions. Moreover, the Bragg modes observed in the PL spectra above 10 kW/cm² can be identified by comparison to those observed on the reflectivity spectrum. As observed for the bulk layer the broadening of the excitonic peaks makes difficult their identification above 1.1 kW/cm² (red spectrum). When increasing the power above 45 kW/cm², numerous narrows lines appear near 3450 meV. This energy corresponds to the electron-hole plasma emission detected as a broad band on the bulk layer at large power. The excitons have ceased to exist as individual quasiparticles. This lasing effect is attributed to stimulated emission arising from the population inversion in the active layer obtained at high excitation power. As a conclusion, the power used to obtain polariton lasing remains far from 1.1 kW/cm², which constitutes the limit below which excitons still exist.

**Caption**

*Figure 1: Far field emission of ZnO MC at RT below (a, c) and at Pth (b, d) for two different detuning values (-98 and -44 meV) corresponding to photonic fractions of 76 and 63% respectively. The calculated dispersion curves are displayed as solid line for the LPB, dash line for the bare cavity dispersion and dash dot lines for the Bragg polariton modes (B1 and B2).*

*Figure 2: Far field emission of GaN MC at RT below (a, c) and at Pth (b, d) for two different detuning values (-83 and -56 meV) corresponding to photonic fractions of 85 and 76.5% respectively. The calculated dispersion curves are displayed as solid line for the LPB, dash line for the bare cavity, and dash dot lines for the Bragg polariton modes (B1, B2 and B3).*

*Figure 3: (a) Typical evolution of LPB (82% photon) at RT on GaN MC which confirms the weak photonic disorder. The same evolution has been observed on the ZnO MC (not shown here). The inset displays a µPL spectrum, a large cavity quality factor equal to 3550 has be determined for the larger photonic weight available. (b) Polariton lasing threshold as a function of the excitation spot diameter for GaN and ZnO MCs at RT.*

*Figure 4: ZnO phase diagram at the intermediate temperature of 100 K. The different regimes of the condensation phase diagram are recalled. The positive (negative) detuning part corresponds to the thermodynamic (kinetic) region. Between these two regimes a trade-off zone can be defined where the polariton lifetime is close to the relaxation time. The vertical dashed lines indicate the detuning corresponding to the energy between 1 or 2 LO phonon resonances and LPB at k=0.*

*Figure 5: GaN phase diagrams (a) at 10, 77, 100 K, (b) 150, 220 and 300 K. The solid arrows lines indicate the detuning corresponding to the energy between 1 LO phonon resonance and LPB at k=0.*

*Figure 6: ZnO phase diagrams (a) at 10, 77, 100 K, (b) 150, 220 and 300 K. The solid arrows lines indicate the detuning corresponding to the energy between 1 or 2 LO phonon resonances and LPB at k=0.*

*Figure 7: Calculated phase diagram based on semi-classical Boltzmann equations for GaN and ZnO MCs at 20, 77 and 300 K.*

*Figure 8: The optimal detuning as a function of the temperature is displayed as black squares in (a) GaN and (b) ZnO MCs. The dark blue zone represents the trade-off region and the light blue area corresponds to the uncertainty of the border estimation. The evolution of the detuning corresponding to the energy resonance between 1 LO (2 LO) phonon resonances and LPB at k=0 is shown as blue (green) solid line.*

*Figure 9: Comparison of polariton lasing thresholds in GaN and ZnO MCs at (a) 10 K versus the polaritonic trap depth and at (b) RT versus the detuning.*

*Figure A1: Photoluminescence spectra recorded for various excitation intensities at 5 K on (a) bulk GaN, (b) same measurements with a zoom on the excitonic transitions. The excitons still exist below 1.1 kW/cm²*

*Figure A2: (a): Photoluminescence spectra recorded for various excitation intensities at 5 K on half GaN MC. (b) Reflectivity and PL on GaN (half cavity) at 5 K. The Bragg modes observed on the PL spectrum are in accordance with the oscillations detected in the reflectivity spectrum.*

*Figure A3: Maximal intensity of the LPB emission of GaN and ZnO MCs versus the optical pumping at RT. In both systems the optimal detuning is chosen and it corresponds to the 2 LO resonance for ZnO.*

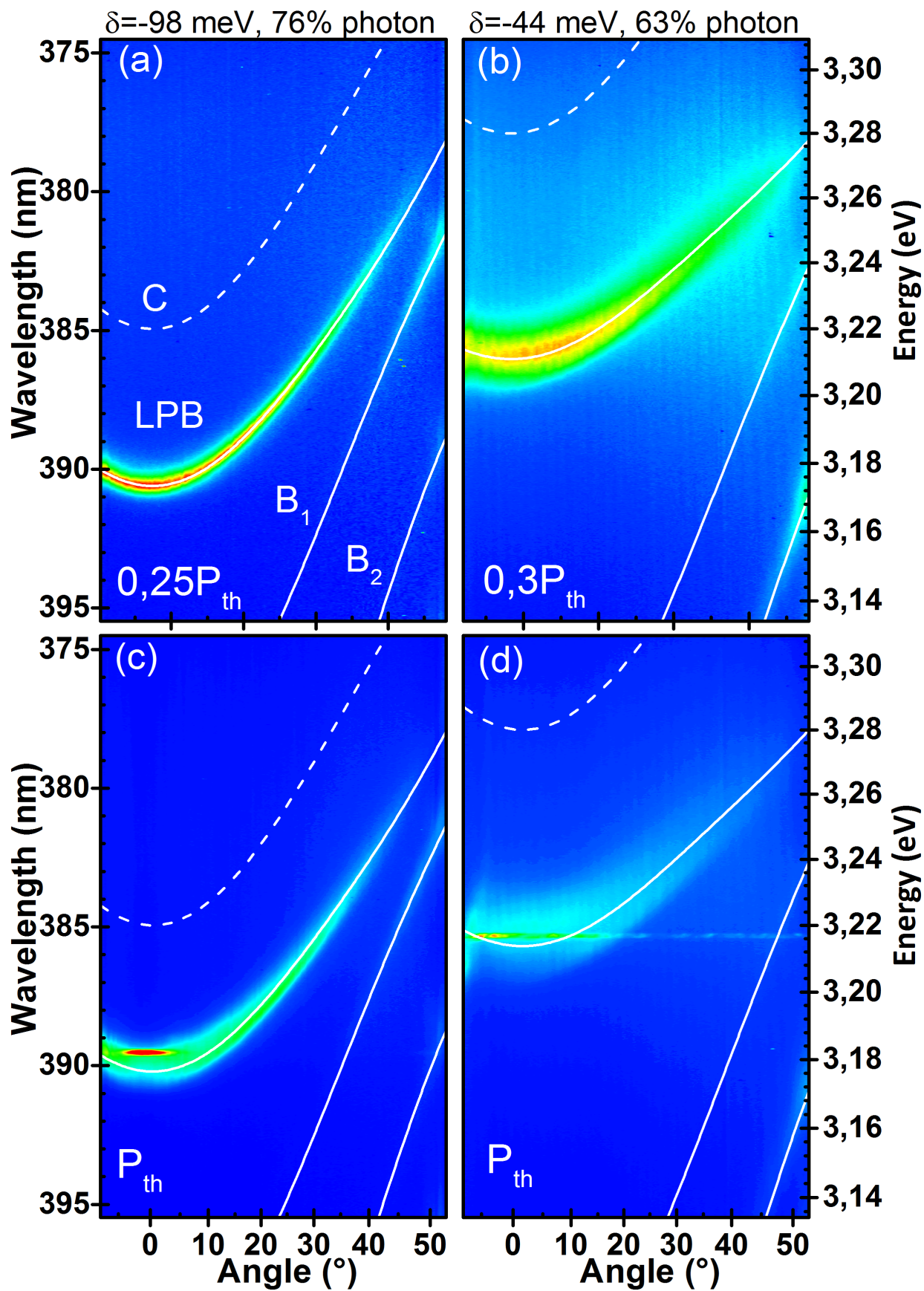

Figure 1

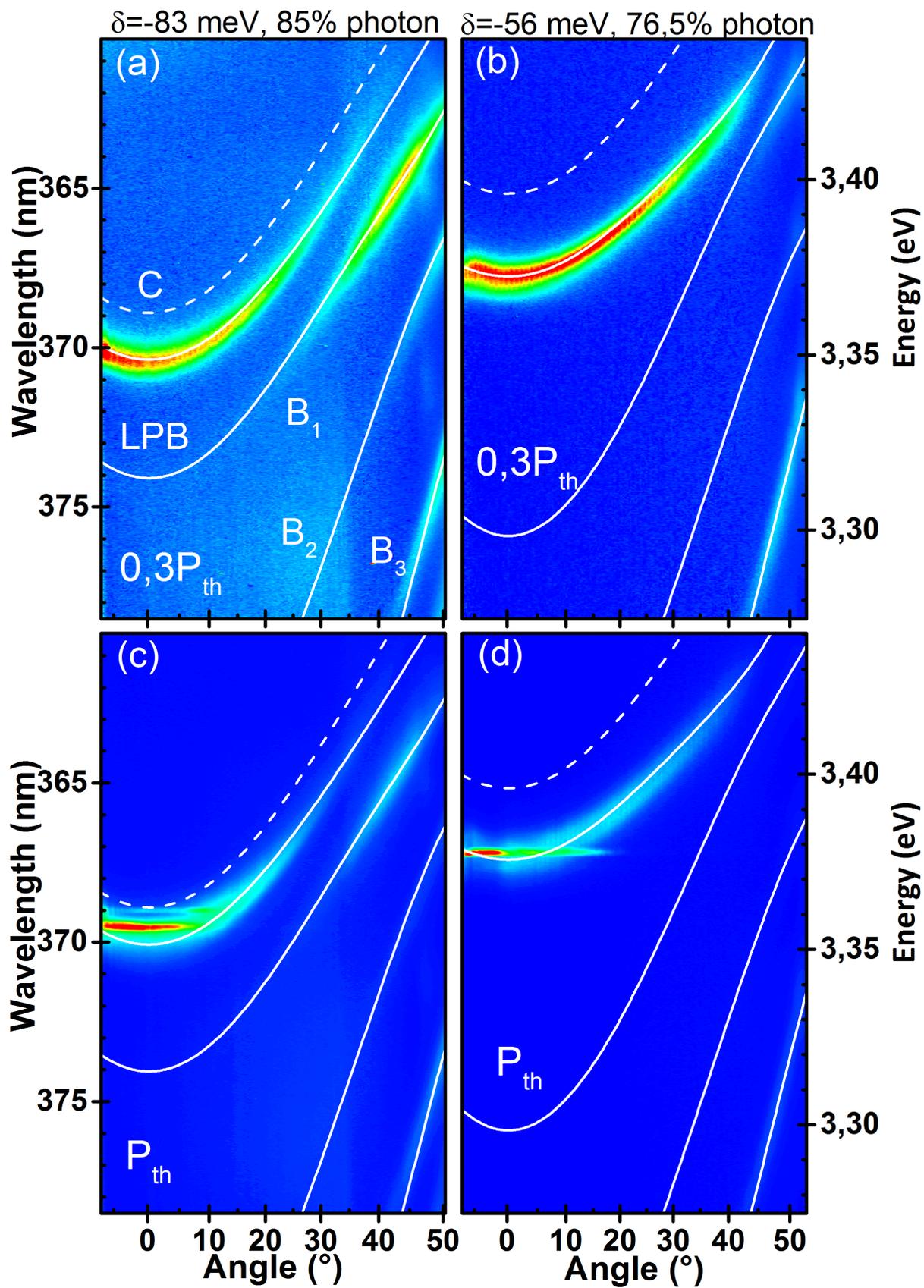

Figure 2

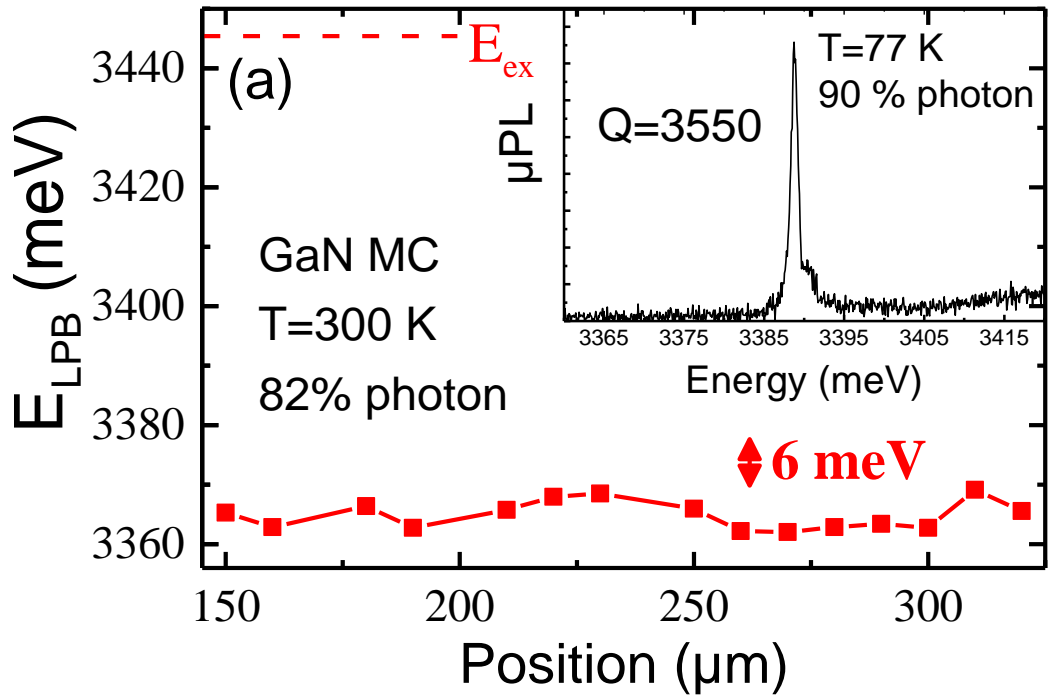
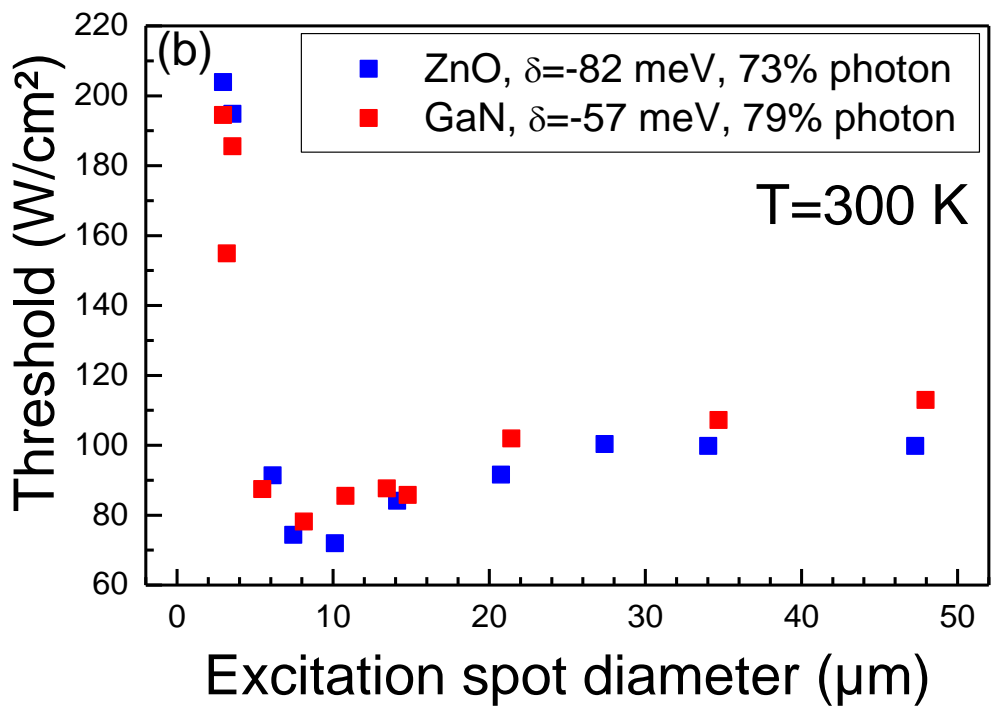

Figure 3

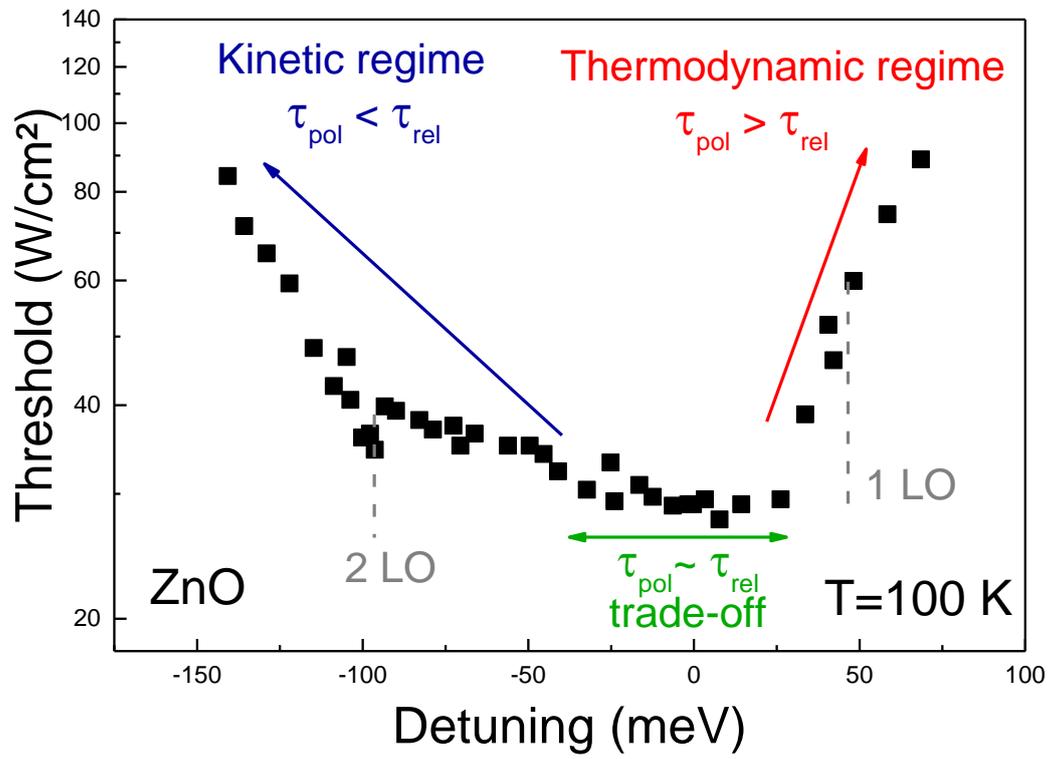

Figure 4

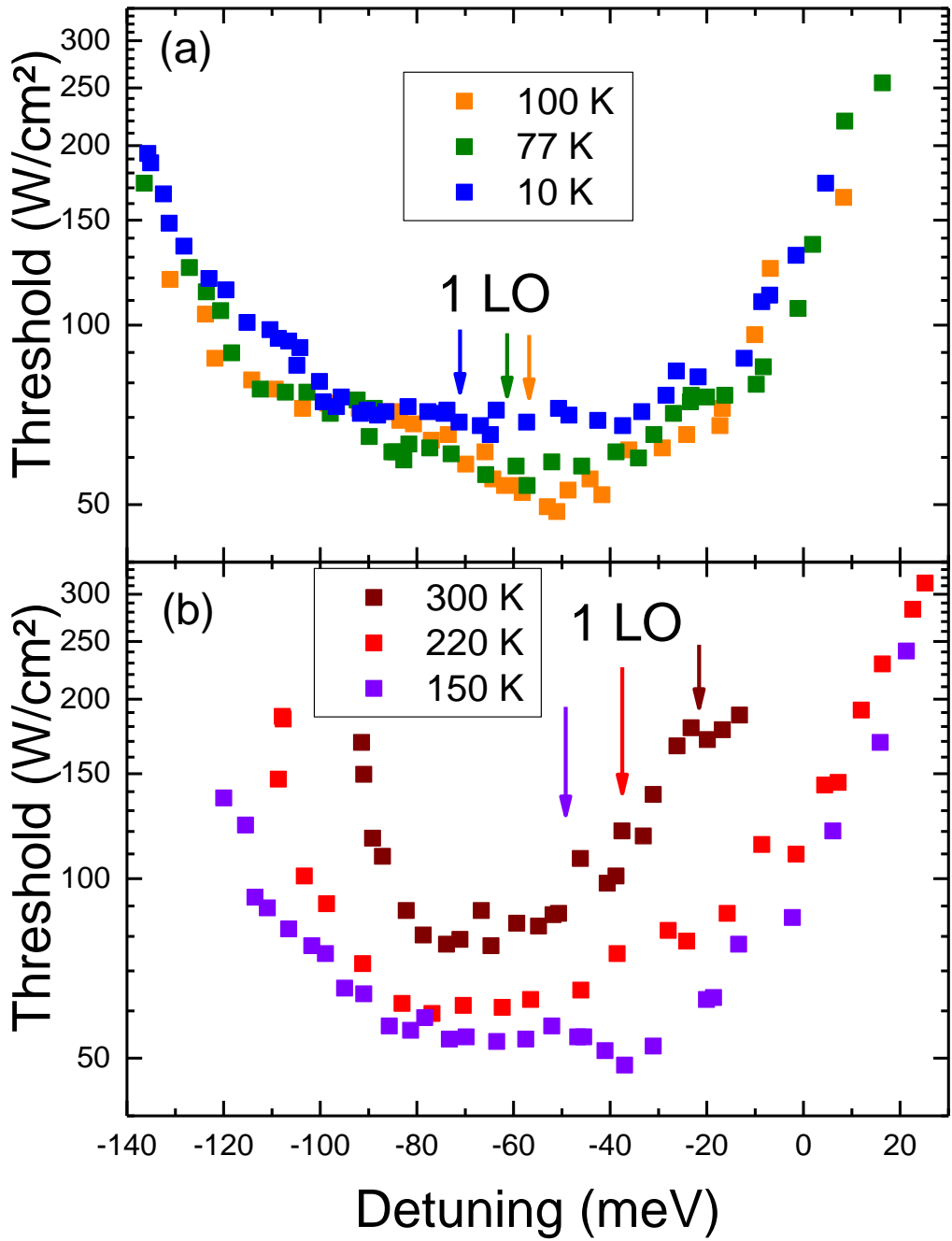

Figure 5

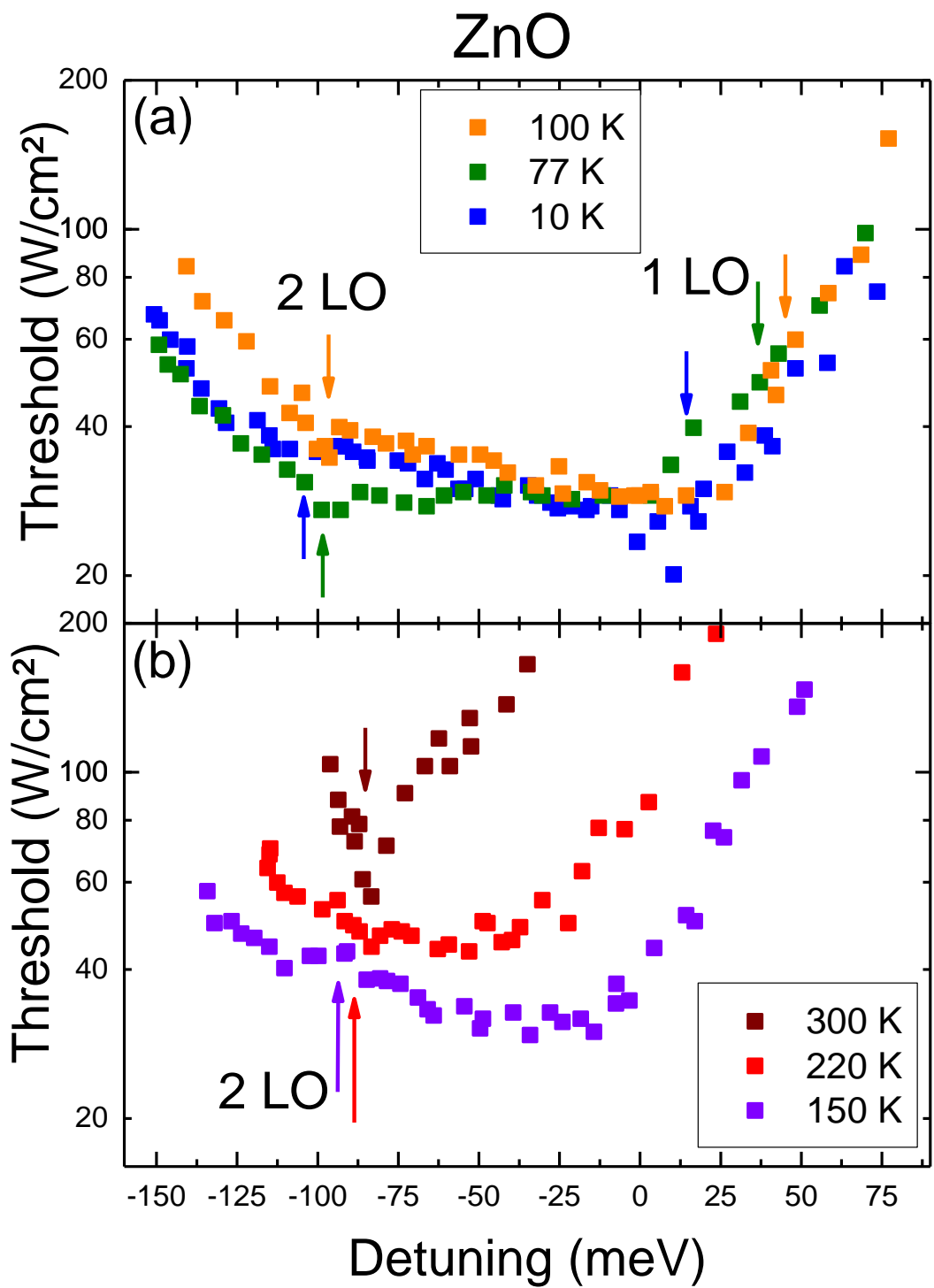

Figure 6

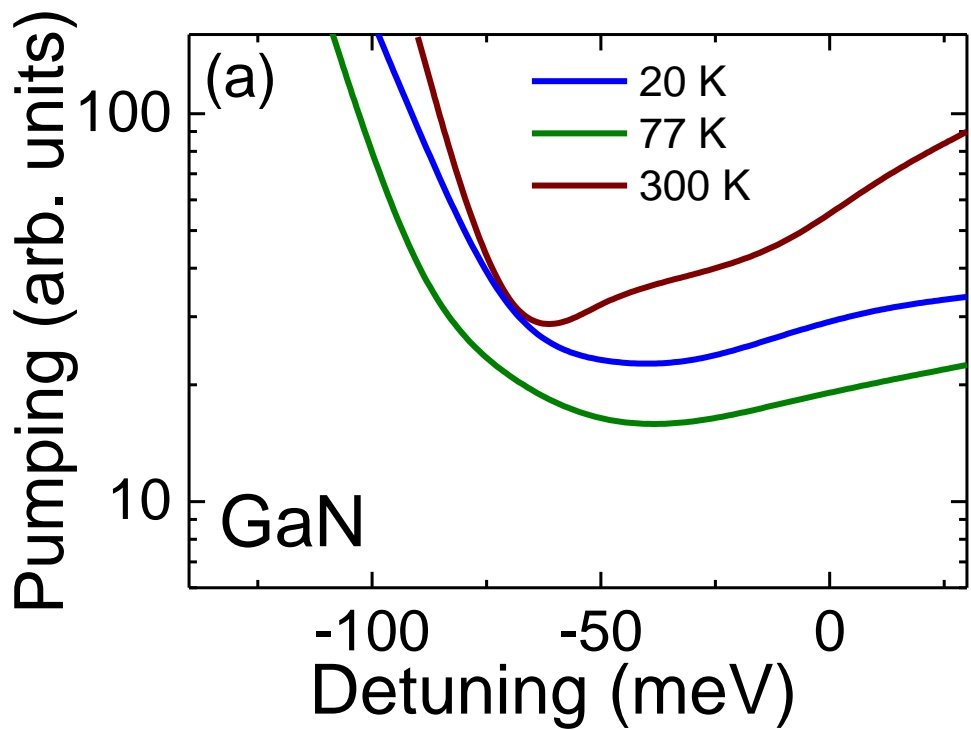
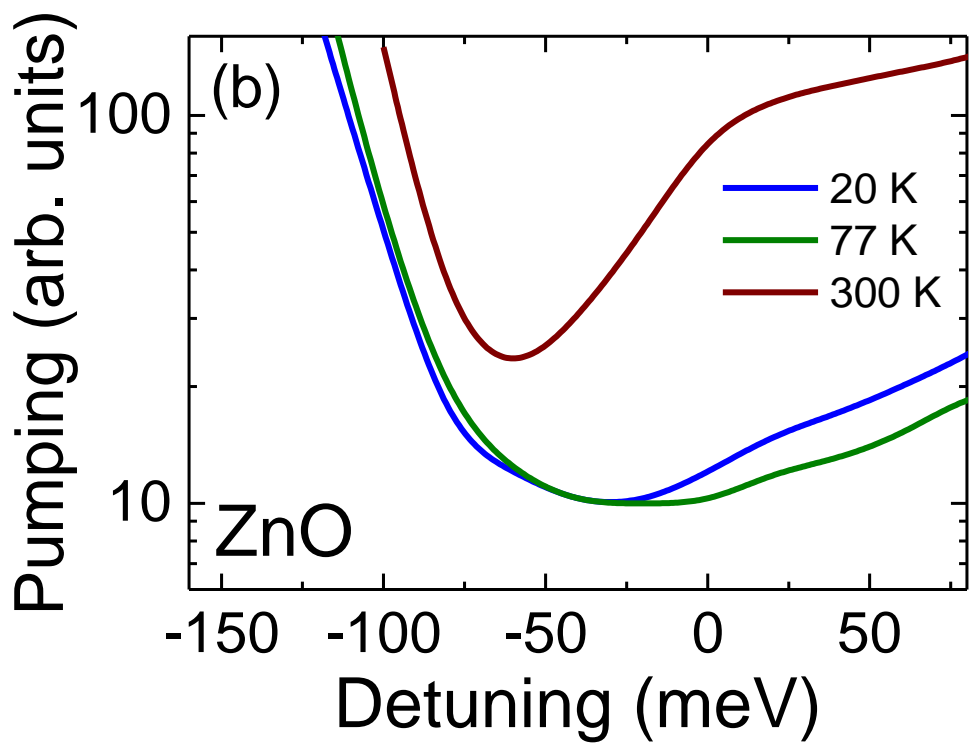

Figure 7

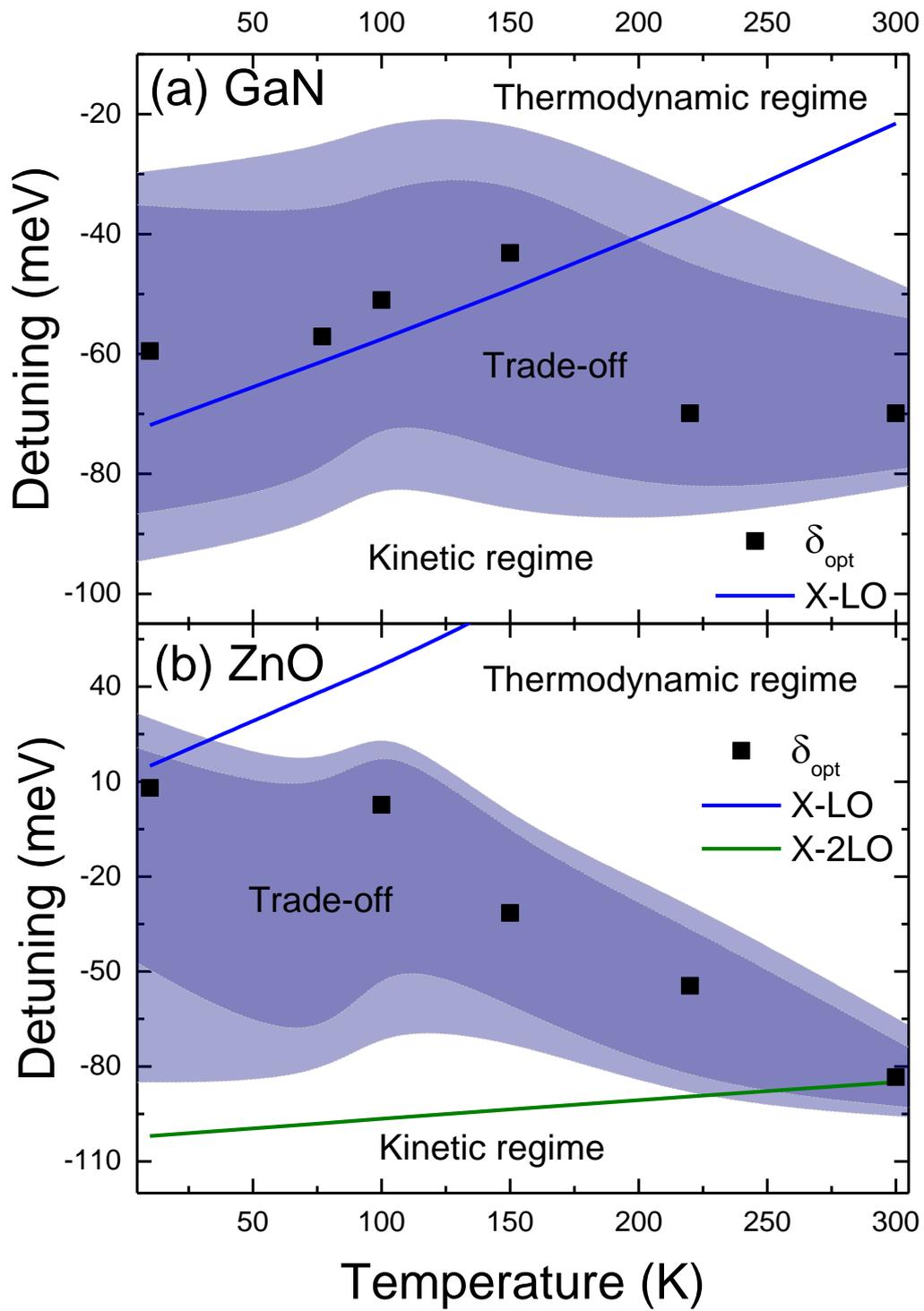

Figure 8

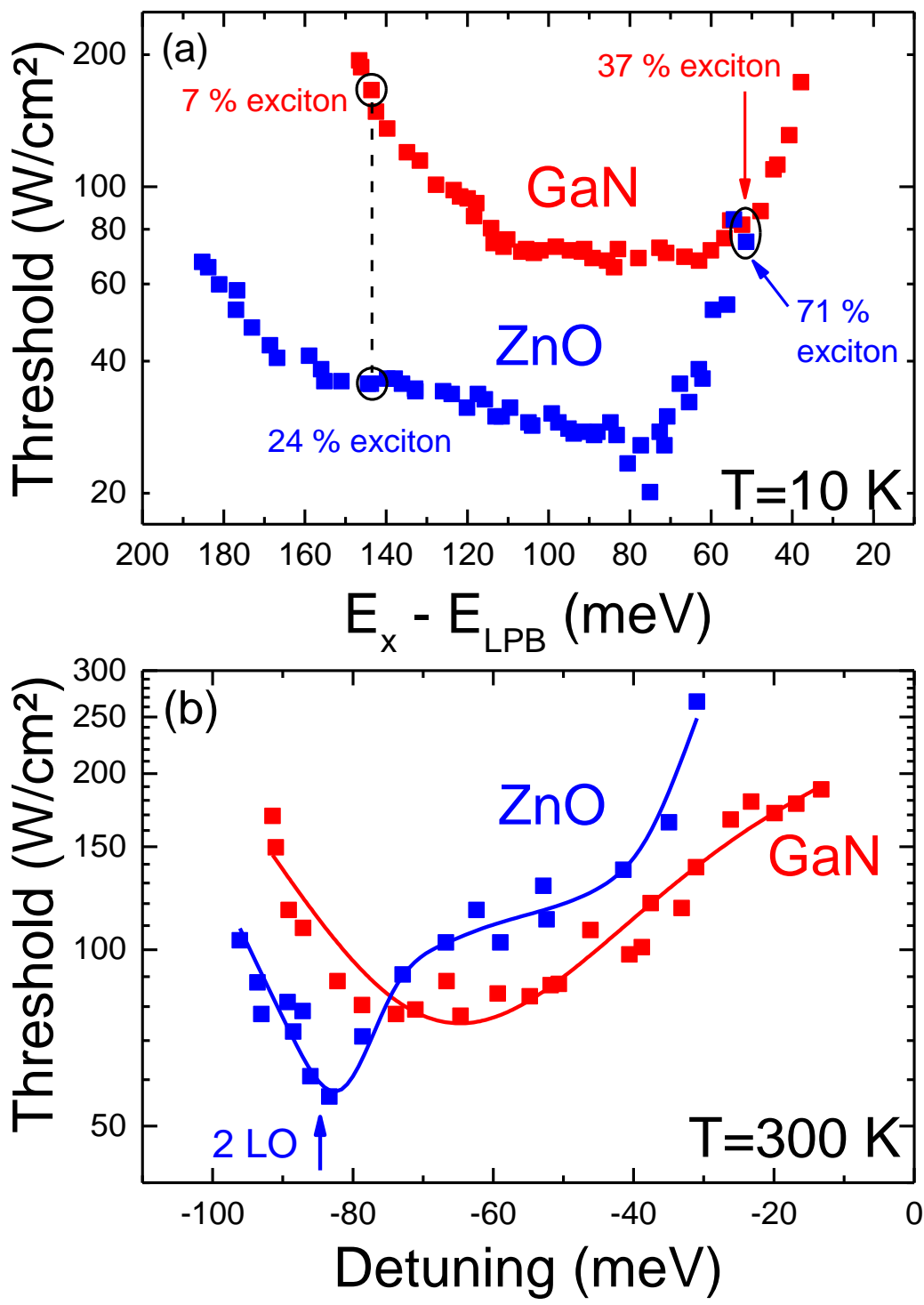

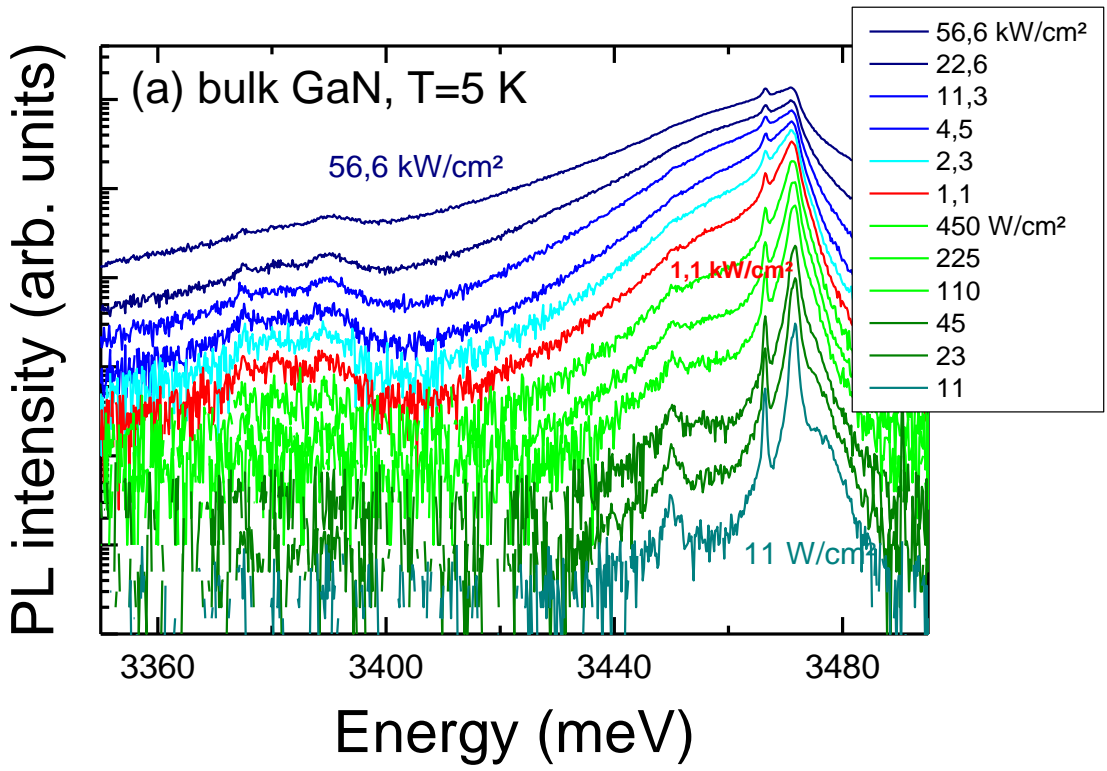

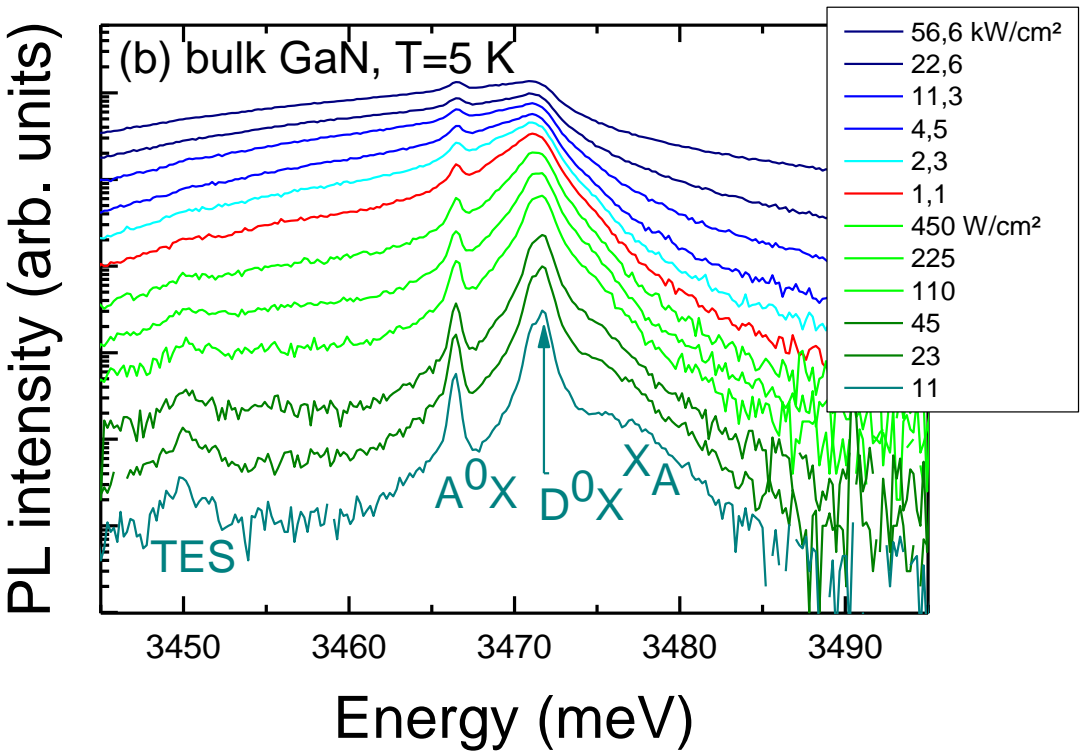

Figure A1

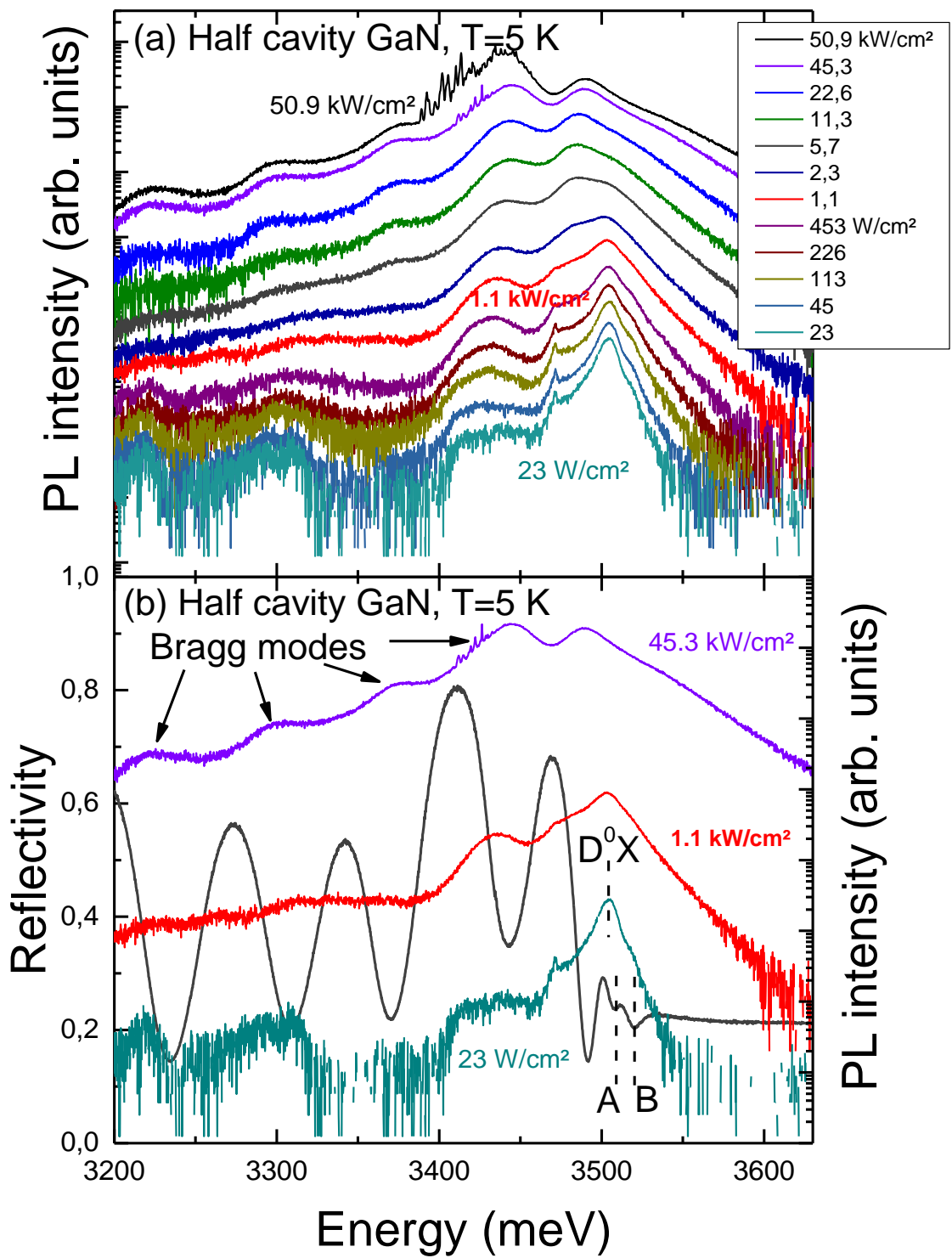

*Figure A2*

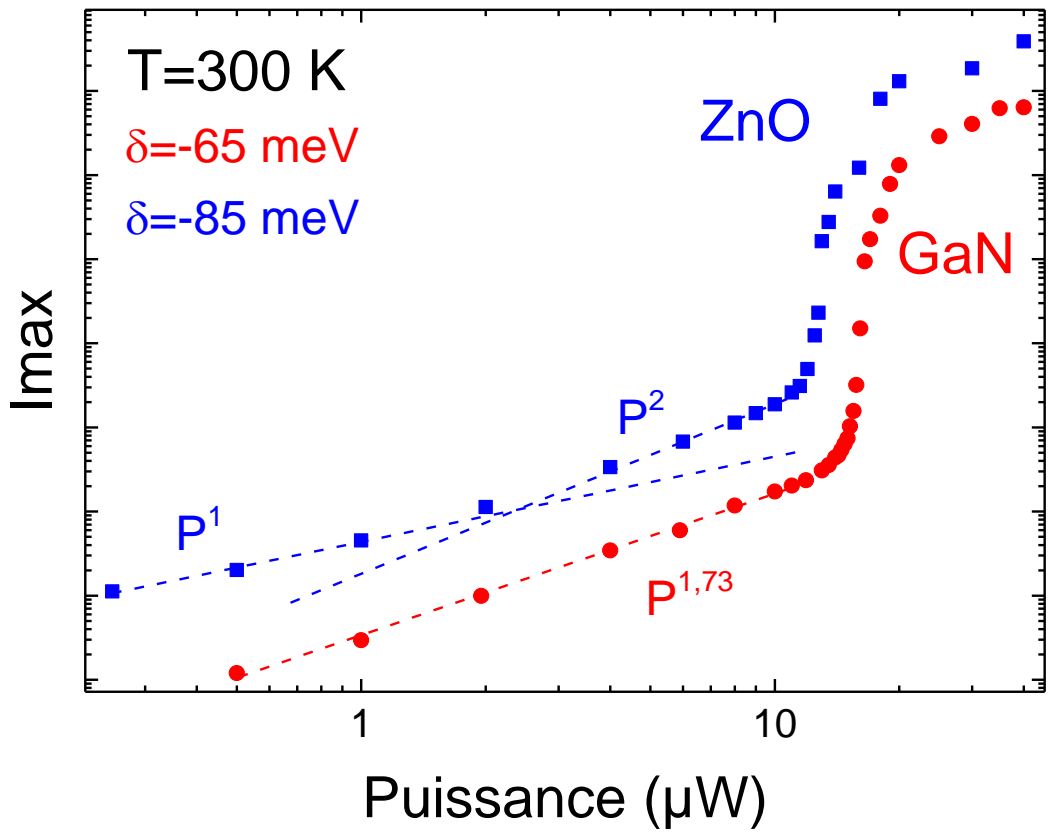

*Figure A3*